\documentclass[conference]{IEEEtran}
\IEEEoverridecommandlockouts
\usepackage{cite}
\usepackage{amsmath,amssymb,amsfonts}
\usepackage{algorithmic}
\usepackage{graphicx}
\usepackage{textcomp}
\usepackage{xcolor}
\usepackage[hyphens]{url}
\usepackage{subcaption}
\usepackage{multirow}
\usepackage[hidelinks,bookmarks=false]{hyperref}
\def\BibTeX{{\rm B\kern-.05em{\sc i\kern-.025em b}\kern-.08em
    T\kern-.1667em\lower.7ex\hbox{E}\kern-.125emX}}
\begin{document}

\title{Self-supervised Light Field View Synthesis Using Cycle Consistency
\thanks{This publication has emanated from research conducted with the financial support of Science Foundation Ireland (SFI) under the Grant Number 15/RP/2776. We also gratefully acknowledge the support of NVIDIA Corporation with the donation of the Titan Xp GPU used for this research.}
}

\author{\IEEEauthorblockN{Yang Chen, Martin Alain, Aljosa Smolic}
\IEEEauthorblockA{V-SENSE project, School of Computer Science and Statistics, Trinity College Dublin \\
\{cheny5, alainm, smolica\}@scss.tcd.ie}
}

\maketitle

\begin{abstract}
High angular resolution is advantageous for practical applications of light fields.
In order to enhance the angular resolution of light fields, view synthesis methods can be utilized to generate dense intermediate views from sparse light field input.
Most successful view synthesis methods are learning-based approaches which require a large amount of training data paired with ground truth.
However, collecting such large datasets for light fields is challenging compared to natural images or videos.
To tackle this problem, we propose a self-supervised light field view synthesis framework with cycle consistency. The proposed method aims to transfer prior knowledge learned from high quality natural video datasets to the light field view synthesis task, which reduces the need for labeled light field data. A cycle consistency constraint is used to build bidirectional mapping enforcing the generated views to be consistent with the input views. 
Derived from this key concept, two loss functions, cycle loss and reconstruction loss, are used to fine-tune the pre-trained model of a state-of-the-art video interpolation method.
The proposed method is evaluated on various datasets to validate its robustness, and results show it not only achieves competitive performance compared to supervised fine-tuning, but also outperforms state-of-the-art light field view synthesis methods, especially when generating multiple intermediate views. Besides, our generic light field view synthesis framework can be adopted to any pre-trained model for advanced video interpolation.
\end{abstract}


\section{Introduction}
Light field imaging has been introduced to the computer graphics and computer vision community over 20 years ago in the pioneer work of Levoy et al.~\cite{Levoy1996}, and has gained a lot of attention since.
Compared to traditional 2D imaging systems, 4D light field imaging systems aim to collect all light rays passing through a given 3D volume by capturing not only two spatial but also two additional angular dimensions.
Light field imaging has many applications ranging from post-capture photograph refocusing to virtual and augmented reality~\cite{yu2017light}. 


Advantages of dense light fields have been demonstrated for various computer vision and graphics tasks compared to sparse light fields, including depth estimation, object segmentation and image-based rendering~\cite{wu2017overview}.
However, a trade-off has usually to be made between spatial and angular resolution when implementing light field capture systems.
While lenslet plenoptic cameras tend to favor angular resolution over spatial resolution~\cite{Ng2005}, modern camera arrays allowing light field video capture have physically limited angular sampling but can capture high resolution 2D views~\cite{SAUCE}.

Thus, light field resolution enhancement has been a very active research topic.
While enhancing the spatial resolution can be seen as an extension of 2D image super-resolution methods~\cite{alain2018light,yoon2015learning},
enhancing the angular resolution has been explored from various angles, such as angular super-resolution, view synthesis or interpolation, light field reconstruction, or epipolar-plane image (EPI) inpainting~\cite{Shearlet,yoon2015learning,kalantari2016learning,LFEPICNN_journal,wang2018end}.
All these methods eventually generate novel intermediate views from a sparse set of input views and thus obtain a denser light field.

The study of these angular resolution enhancement approaches shows that learning-based methods achieve the best performance, and motivates the work presented in this paper. 
One of the first learning-based approaches for light field view synthesis was introduced in~\cite{kalantari2016learning}, and proposed to train two convolutional neural networks (CNN) first estimating the disparity, and then using this disparity to synthesize the intermediate views. It has been observed that the method proposed by Kalantari et.al.~\cite{kalantari2016learning} can fail when dealing with complicated scenes containing occluded regions, non-Lambertian areas and large displacements.
A "blur-detail restoration-deblur" framework was then proposed to enhance light field angular resolution using EPIs in~\cite{LFEPICNN_journal}.
An end-to-end network with 3D detail on pseudo 4D EPI was then introduced in~\cite{wang2018end}.
These generative methods are prone to over-smoothing for large displacements that leads to a loss in perceptual quality.
The limitations of these learning-based methods can partly be explained by the limited size of the training datasets.

As mentioned above, implementing real-world light field capture systems is a complex task, which inherently limits the size of available datasets, and can only be partly compensated by the use of synthetic datasets.
In comparison, a huge amount of real-world natural images and videos are readily available, which allows to train powerful learning-based methods, and it has been shown recently that a direct extension of single image learning-based methods can outperform light field learning-based approaches for spatial super-resolution~\cite{cheng2019light}.

To address the aforementioned issues, we propose a self-supervised learning-based light field view synthesis framework based on existing video view synthesis methods which benefit from large real-world training datasets.

Rather than training a leaning-based method from scratch, the proposed framework adapts the existing video view synthesis method to light fields using fine-tuning only, which requires smaller training datasets.
Cycle consistency has proven capable of modeling invertible mapping when direct supervision is unavailable~\cite{zhu2017unpaired,liu2019deep,reda2019unsupervised}.
Thus, a cycle consistency constraint is introduced in our framework to allow self-supervised training without the need of paired ground truth, which further reduces the size of training data required.

In summary, our main contributions in this paper are:
\begin{itemize}
    \item We introduce a novel light field view synthesis framework utilizing natural priors from existing large image datasets.
    \item We propose for the first time a self-supervised fine-tuning training approach for light fields based on the cycle consistency constraint.
    \item We demonstrate that the proposed framework outperforms state-of-the-art light field view synthesis methods. 
\end{itemize}

This paper is organized as follows. In Section~\ref{sec:related}, we review existing related work about light field angular super-resolution, general video frame interpolation and applications of the cycle consistency. In Section~\ref{sec:method}, the proposed light field view synthesis framework and related self-supervised training details are explained. Then, our proposed method is evaluated on various datasets and compared to state-of-the-art light field view synthesis methods in Section~\ref{sec:results}. Finally, we present our conclusions in Section~\ref{sec:conclusion}.

\section{Related Work}
\label{sec:related}

\subsection{Light Field View Synthesis}

\subsubsection{Optimization-based methods.}
Dense light fields are very sparse in the transform domain, which can be used as a powerful prior.
Therefore, Shi et al. proposed an optimization framework in the continuous Fourier domain to reconstruct the dense light field~\cite{shi2014light}.
A more advanced framework using the shearlet transform is proposed in~\cite{Shearlet} which performs inpainting on the EPIs to recover the missing views.
While these methods can achieve competitive performance, they require specific inputs which are not flexible and can make them difficult to use in practice.

\subsubsection{Learning-based methods.}

Kalantari et al. proposed a learning based framework for light field view synthesis~\cite{kalantari2016learning} which estimates with a first CNN the disparity from 4 corner input views of a dense light field.
A second CNN is then used to synthesize the target intermediate views using both the disparity and the 4 input views.
Wu et al. re-modeled light field angular super-resolution as a detail restoration problem in the 2D EPI space~\cite{LFEPICNN_journal}.
A detail restoration framework is built to process EPIs of a sparse light field and to recover the angular details with a CNN.
To exploit the inherent consistency of the light field, Wang et al. introduced an end-to-end network with pseudo 4D convolution by combining a 2D convolution on EPIs and a sequential 3D convolution~\cite{wang2018end}.
Yeung et al.~\cite{wing2018fast} also propose a two-step method, which first generates the whole set of novel views using a view synthesis network, and then retrieves texture details using a view refinement network.

Zhou et al.~\cite{zhou2018stereo} train a deep network to predict the Multi-Plane Images (MPI) representation from a narrow-baseline stereo image pair.
The MPI representation can be use to generate novel views using homographies to reproject each plane of the MPI to the desired viewport.
Mildenhall et al.~\cite{mildenhall2019local} extend this method to light fields by promoting an MPI to each view of the light field.
Novel views are then synthesized by combining intermediate generated views from the closest MPIs.

Video interpolation has also been applied to light field view synthesis in~\cite{gao2018parallax}, using fully supervised fine-tuning with conventional loss functions on a small light field dataset. 
Although various existing methods have been proposed to enhance the angular resolution with impressive results, the robustness and generalization of these methods are limited by the quantity and quality of available real-world light field datasets, as explained in the previous section.

\subsection{Video Frame Interpolation}

Niklaus et al. proposed to estimate motion and color interpolation within one stage using an adaptive 2D kernel which is estimated from a trained convolutional neural network~\cite{AdaConv}.
However, 2D kernel estimation requires huge memory to store information for all pixels and this shortcoming is addressed by replacing the 2D kernel with two separable 1D convolutional kernels~\cite{SepConv}.

\subsection{Cycle Consistency}

The key element of our proposed method is the introduction of the cycle consistency to the light field angular domain. The cycle consistency constraint aims to regularize structured predictions and establishes bidirectional mapping instead of unidirectional mapping built by conventional cost measurement functions.


Our work is inspired by the success of cyclic image generation for video interpolation~\cite{liu2019deep,reda2019unsupervised}, which demonstrated the strength of cycle consistency to adapt a pre-trained model to a new target domain. 

\begin{figure}[!t]
\centering
\includegraphics[width=\linewidth]{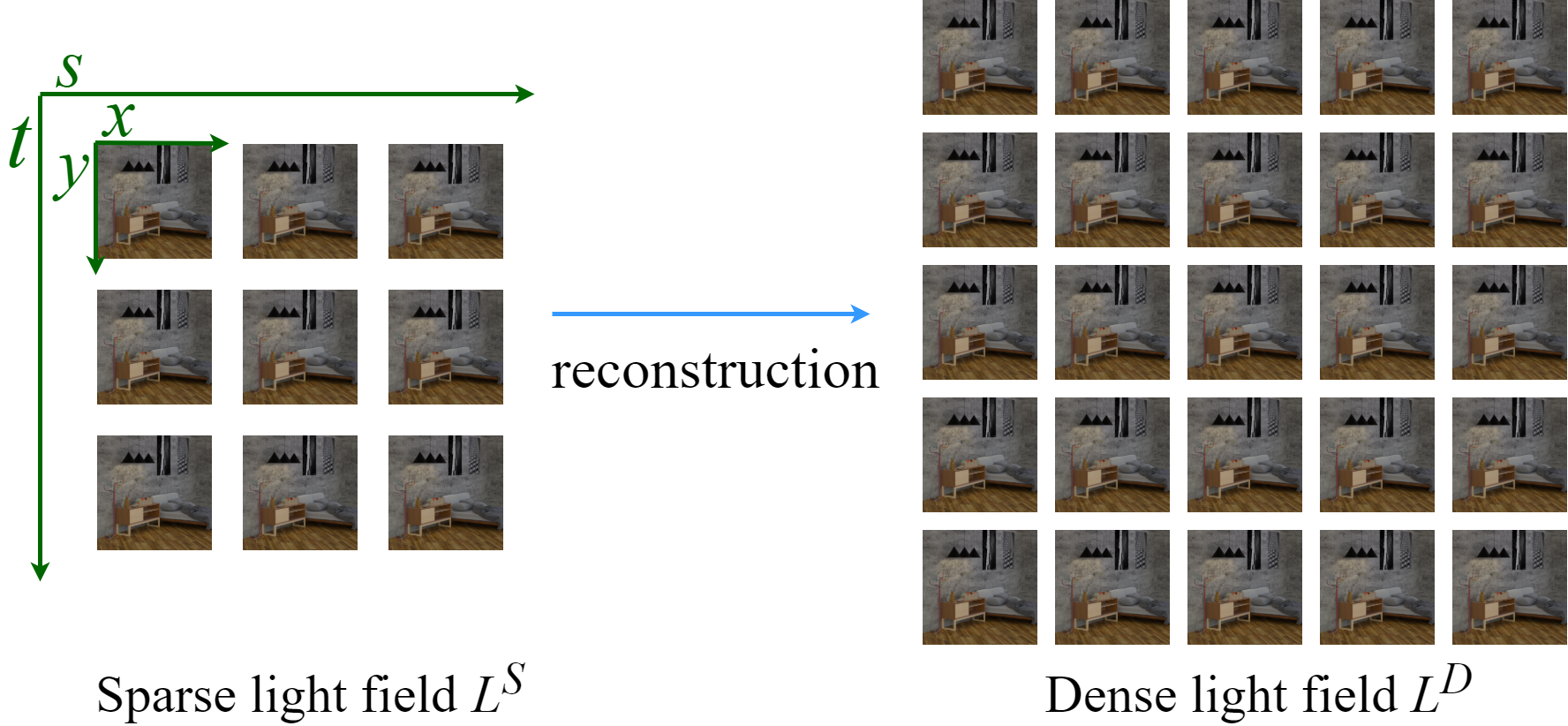}
\caption{\small{Dense light field reconstruction. We aim to reconstruct a dense light field $L^{D}$ with angular resolution $(N, N)$ from a sparse light field $L^{S}$ with angular resolution $(n, n)$. The spatial resolution $(h, w)$ of each view remains unchanged.}}
\label{fig:LF_intro}
\vspace{-0.5cm}
\end{figure}

\section{Light Field View Synthesis Using Cycle Consistency}
\label{sec:method}

\subsection{Problem Formulation}
In this work, a 4D light field $L$ is parameterized using the two parallel planes representation as depicted in Figure~\ref{fig:LF_intro}, indexed by $x$, $y$ over the spatial dimensions and $s$, $t$ are the angular dimensions.
We denote by $I_{s,t}$ the view extracted from a light field $L$ at the angular position $s,t$.
Given a sparsely-sampled light field $L^{S}$ with resolution  $(h \times w \times n \times n)$, the goal is to reconstruct a more densely-sampled light field $L^{D}$ with the same spatial resolution and a higher angular resolution $(h \times w \times N \times N)$, where $N = \alpha (n - 1) + 1$ and $\alpha$ is the up-sampling factor in the angular domain. Unless mentioned specifically, $\alpha = 2$ is used as default to explain our method. 
By fixing one angular dimension, a set of views can be extracted along the remaining angular dimension of the light field.
Such a view set can be considered as a consecutive frame sequence, which can be captured by a virtual camera moving along the corresponded angular direction.
Thus, the dense light field reconstruction problem can be treated as a video interpolation process along the fixed angular dimension.
As listed in section \ref{sec:related}, many CNN-based methods have been shown to be successful for video interpolation tasks.
However, directly adopting a pre-trained network from an existing  video interpolation method to the light field domain may fail since the distribution of these two kind of data may differ.
On the other hand, retraining a CNN from scratch can be laborious and the limited size of light field datasets may not allow to reach competitive performance.
Thus, to maximally leverage the advantage of the cutting-edge video interpolation methods and to avoid its troublesome retraining, we introduce a self-supervised fine-tuning approach using cycle consistency to apply the pre-trained model of a video interpolation method to the light field domain.

\subsection{Proposed Framework with Self-Supervised Learning}

\begin{figure}[!t]
\centering
\includegraphics[width=\linewidth]{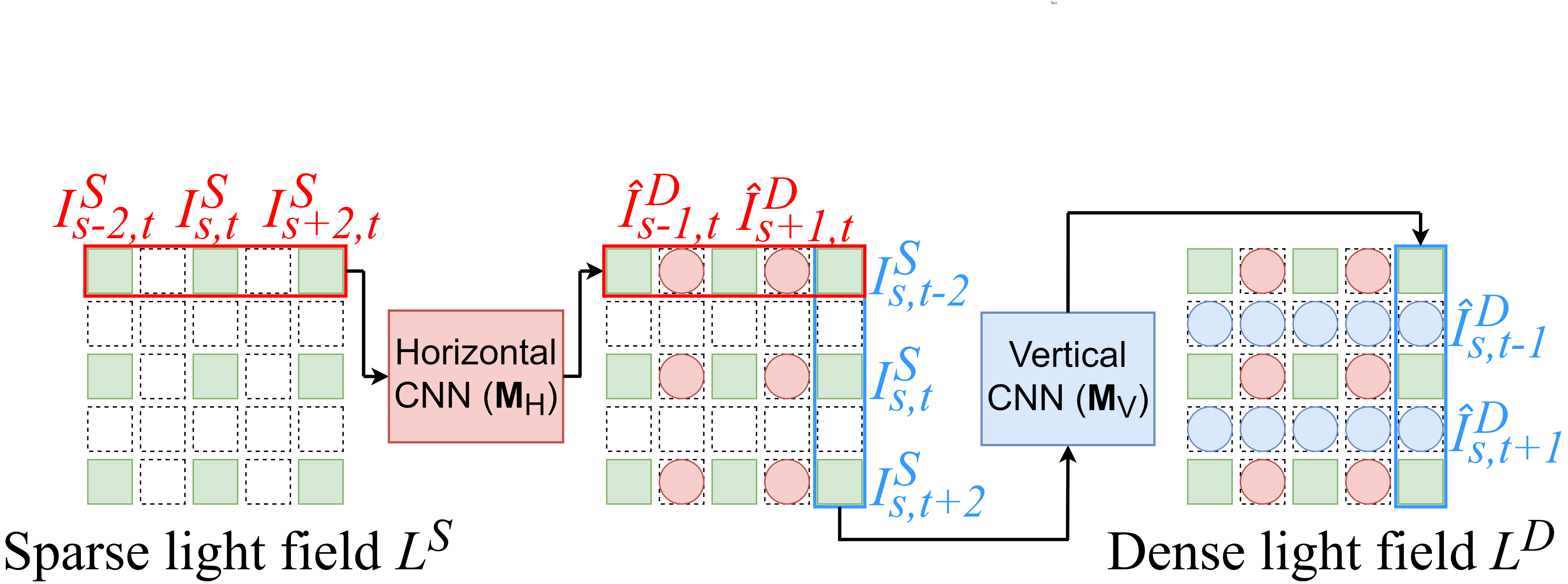}
\caption{\small{An overview of our proposed view interpolation approach. Horizontal and vertical interpolation are cascaded to reconstruct $L^{D}$ from $L^{S}$ using view triplets from the corresponding angular dimensions.}}
\label{fig:overview}
\vspace{-0.3cm}
\end{figure}

Given a sparse light field $L^{S}$ with angular resolution $(n, n)$, our proposed approach aims to build a learning based model that takes this light field as input and accurately reconstructs a high quality dense light field $L^{D}$ with angular resolution $(N, N)$ without the support of paired ground truth.
As shown in Figure~\ref{fig:overview}, we consider triplets of views extracted from $L^{S}$ along a fixed angular dimension, either horizontally $\{I^S_{s-2,t},I^S_{s,t},I^S_{s+2,t}\}$ or vertically $\{I^S_{s,t-2},I^S_{s,t},I^S_{s,t+2}\}$.
Note that triplets have to be used due to our proposed cycle loss described below.
A dense light field $L^{D}$ is obtained by first performing horizontal interpolation on all rows, and then performing vertical interpolation on all columns.
The view interpolation is achieved by two CNNs which share the same architecture but are trained separately along the horizontal and vertical dimensions.

Let us consider the horizontal interpolation case in order to explain the framework more in detail.
Given an input triplet $\{I^S_{s-2,t},I^S_{s,t},I^S_{s+2,t}\}$, two intermediate views can be generated from pairwise adjacent views: 
\begin{equation}
\label{eq:triplet_interp}
\begin{aligned}
  \hat{I}^D_{s-1,t} &= \textbf{M}(I^S_{s-2,t},I^S_{s,t})  \\
  \hat{I}^D_{s+1,t} &= \textbf{M}(I^S_{s,t},I^S_{s+2,t})
\end{aligned}
\end{equation}
\noindent where \textbf{M} is a pre-trained video interpolation method.

\begin{figure}[t]
    \begin{subfigure}{0.23\textwidth}
        \centering
        \includegraphics[width=\linewidth]{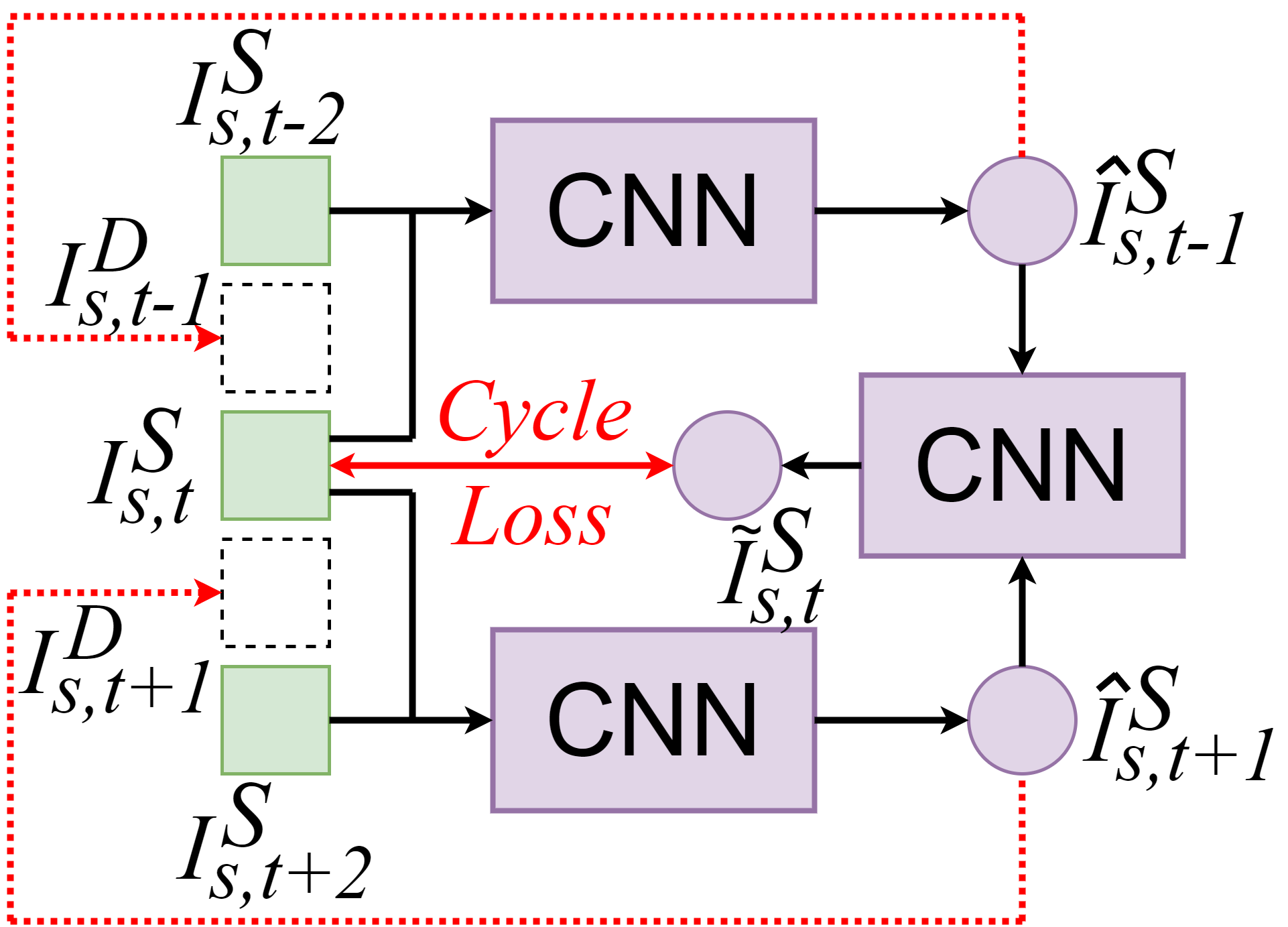} 
        \caption{\small{Cycle loss}}
        \label{fig:fullLF_subimg1}
    \end{subfigure}
    \hfill
    \begin{subfigure}{0.23\textwidth}
        \centering
        
        \includegraphics[width=\linewidth]{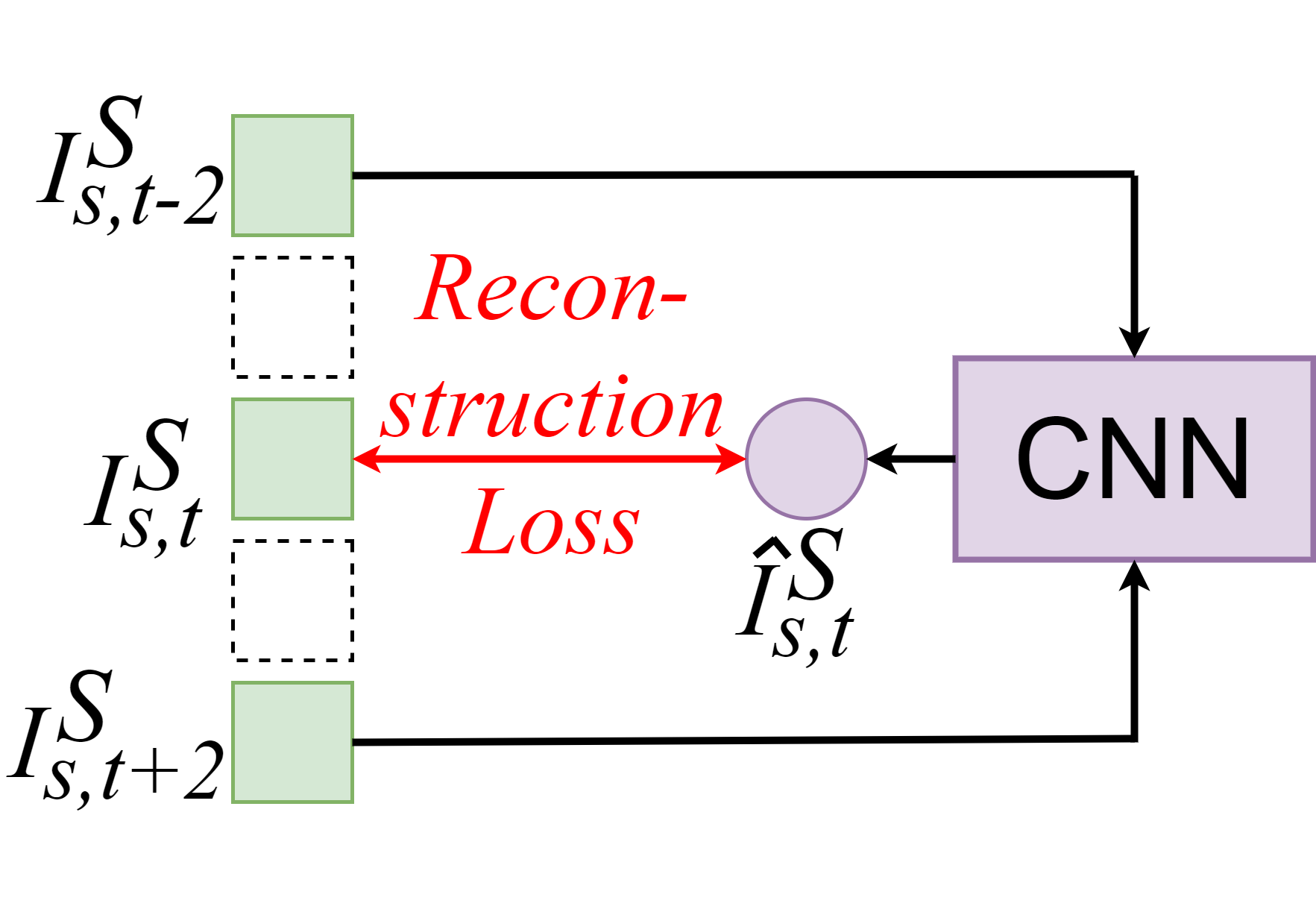}
        \caption{\small{Reconstruction loss}}
        \label{fig:fullLF_subimg2}
    \end{subfigure}
     
    \caption{\small{Illustration of the cycle loss and reconstruction loss on a vertical input triplet. Both losses do not require any knowledge of the ground truth intermediate views (represented by dashed square) and can therefore be used for self-supervised training.}}
    \label{fig:losses}
    \vspace{-0.5cm}
\end{figure}

Inspired by the recent success of the application of the cycle consistency for video interpolation~\cite{liu2019deep,reda2019unsupervised}, we propose to fine-tune our baseline interpolator \textbf{M} in a self-supervised manner by applying the cycle consistency constraint to the light field angular domain, as shown in Figure~\ref{fig:fullLF_subimg1}. 
By applying the interpolator \textbf{M} on the two intermediate views generated from the input triplet as defined in equation \ref{eq:triplet_interp}, we can obtain an estimate of the center view of the input triplet $I^S_{s,t}$ which we denote as the cycle-reconstructed view $\Tilde{I}^S_{s,t}$:

\begin{equation}
\begin{aligned}
    \Tilde{I}^S_{s,t} &= \textbf{M} \left ( \hat{I}^D_{s-1,t},\  \hat{I}^D_{s+1,t} \right ) \\
    &= \textbf{M} \left ( \textbf{M}(I^S_{s-2,t},I^S_{s,t}),\ \textbf{M}(I^S_{s,t},I^S_{s+2,t}) \right )
\end{aligned}
\end{equation}

We can thus define the cycle-loss as the $\ell_{1}$-norm distance between the cycle-reconstructed view $\hat{I}^S_{s,t}$ and the input view $I^S_{s,t}$:

\begin{equation}
    \mathcal{L}_{c} = ||\Tilde{I}^S_{s,t} - I^S_{s,t}||_{1}
    \label{eq:cycleloss}
\end{equation}

\noindent While $\ell_{1}$-norm based losses are able to minimize the overall error between the estimated images and the corresponding original images, they are known to generate over-smooth results.
To tackle this problem, we also introduce in our framework a perceptual loss $\mathcal{L}_{p}$, defined as the $\ell_{2}$-norm between high-level convolutional features extracted from the cycle-reconstructed view and the input view: 
\begin{equation}
    \mathcal{L}_{p} = ||\Psi(\Tilde{I}^S_{s,t}) - \Psi(I^S_{s,t})||_{2}
    \label{eq:vggloss}
\end{equation}
\noindent where $\Psi$ extracts  the  convolutional  features  from  images using  a  VGG-16  network~\cite{simonyan2014very}, which then is applied to train our base CNN network (SepConv~\cite{SepConv}, see below). 

Furthermore, to stabilize the training process, we introduce a reconstruction loss $\mathcal{L}_{r}$, as shown in Figure~\ref{fig:fullLF_subimg2}.
In this case, the two non-adjacent views from the input triplet $I^S_{s-2,t}$ and $I^S_{s+2,t}$ are used to generate the center view of the input triplet:

\begin{equation}
    \hat{I}^S_{s,t} = \textbf{M} \left ( I^S_{s-2,t},  I^S_{s+2,t} \right )
\end{equation}

\noindent This reconstructed view can be used to define the reconstruction loss $\mathcal{L}_{r}$ as its $\ell_{1}$-norm distance to the input view:

\begin{equation}
    \mathcal{L}_{r} = ||\hat{I}^S_{s,t} - I^S_{s,t}||_{1}
    \label{eq:restoreloss}
\end{equation}

Note that all losses introduced in our framework as defined in equations \ref{eq:cycleloss}, \ref{eq:vggloss}, and \ref{eq:restoreloss}, do not rely on any knowledge of the ground truth dense light field $L_D$ but only the given sparse input light field $L_S$, thus allowing to perform self-supervised training or fine-tuning of the learning-based interpolator \textbf{M}.

\begin{figure}[!t]
\centering
\includegraphics[width=0.9\linewidth]{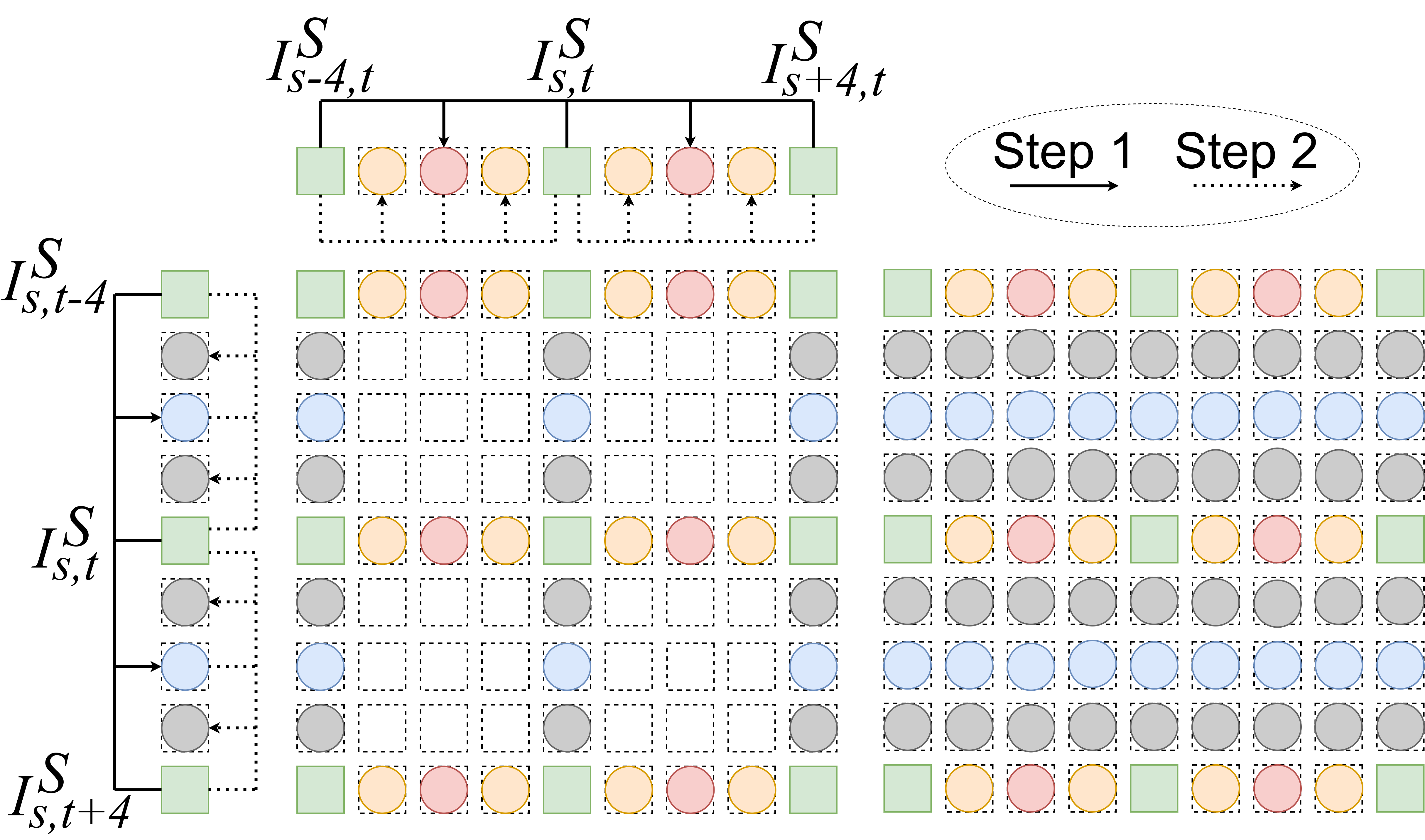}
\caption{\small{Demonstration of the two-step strategy to generate multiple intermediate views from a set of sparse views, denoted as green squares, when $\alpha = 4$. The first step is to synthesize middle views, denoted as red and blue circles, between each pairwise input views. The second step uses original and synthetic views to reconstruct the remaining missing views, denoted as yellow and grey circles, along one angular dimension.}}
\label{fig:MVG}
\vspace{-0.5cm}
\end{figure}

\textbf{Multi-step Light Field Generation}
While our framework naturally performs angular up-sampling with a factor $\alpha = 2$, denser light fields can be obtained by iteratively applying the proposed approach, as illustrated in Figure ~\ref{fig:MVG} for $\alpha = 4$.
Any upsampling factor which is a power of two is in fact supported, \emph{i.e.} $\alpha = 2^{x}, (x \in \mathbb{Z}\ \&\ x > 1)$. 

\subsection{Implementation Details}

In this work, we select the adaptive separable convolution (SepConv)~\cite{SepConv} as our baseline interpolator \textbf{M} due to its balance between ease of use and performance accuracy, but note that any learning-based video interpolation method~\cite{Niklaus2018ContextAwareSF,jiang2018super,DAIN} can be used within our framework.
The network of SepConv employs an encoder-decoder architecture, each part contains convolution blocks and skip connections, to extract features and then performs four 1D kernel estimations individually to obtain the final results.
We use the implementation available online based on \emph{PyTorch}~\footnote{\href{https://github.com/sniklaus/sepconv-slomo}{github.com/sniklaus/sepconv-slomo}}\footnote{\href{https://github.com/HyeongminLEE/pytorch-sepconv}{github.com/HyeongminLEE/pytorch-sepconv}} and use the default configurations from the original SepConv paper.
We fine-tune the pre-trained model by minimizing the objective function:
\begin{equation}
    \underset{\textbf{M}}{\arg \min} \left (  \lambda_{c}\mathcal{L}_{c} + \lambda_{r}\mathcal{L}_{r} + \lambda_{p}\mathcal{L}_{p} \right )
\end{equation}
\noindent where $\mathcal{L}_{c}$, $\mathcal{L}_{r}$ and $\mathcal{L}_{p}$ are defined in equations~\eqref{eq:cycleloss},~\eqref{eq:restoreloss} and~\eqref{eq:vggloss}.
For all experiments, we set the parameters as $\lambda_{c}=1$, $\lambda_{r}=1$, and $\lambda_{p}=0.06$.
The Adam optimizer is applied for optimization with a batch size of 8.
We start with the learning rate of 0.001 and a scheduler is applied to decay the rate according to the learning progress.
As in the original SepConv work, we firstly crop training data to $150 \times 150$ patches, then randomly crop to $128 \times 128$.
In addition, we perform pre-processing to eliminate patches containing too small disparity.
An Intel Core i7-6700k 4.0GHz CPU was used for all our experiments, and the neural network training was run on a single Nvidia Titan Xp GPU with 12 GB memory.

\section{Experiments}
\label{sec:results}

In this section, we first conduct an ablation study, especially evaluating the efficiency of the proposed framework compared to supervised fine-tuning.
For this purpose, we use a variety of real-world and synthetic dense light field datasets which we sub-sample to create our test sparse datasets with sampling ratios $\alpha=2$ and $\alpha=4$.

We then compare the proposed framework to two top-performing state-of-the-art light field view synthesis methods, a shearlet-based  method~\cite{Shearlet} and a learning-based method (LFEPICNN)~\cite{LFEPICNN_journal}.

For all our evaluations, the peak signal-to-noise ratio (PSNR) and the structural similarity (SSIM) are computed over RGB images to evaluate the numerical performance of the different methods.
For each light field, unless emphasized specifically, the average numerical results are computed over all synthesized views. 
All evaluations are performed on the same machine to ensure fairness of the comparison.
Please find more detailed experimental summary on our website~\footnote{\url{https://v-sense.scss.tcd.ie/?p=5163}}.

\subsection{Ablation Study}

For this study we used dense light fields from real-world and synthetic datasets.
For the real-world dataset, we selected 27 real-world Lytro light fields captured by EPFL~\cite{Rerabek2016} and INRIA~\cite{INRIADataset} using Lytro Illum cameras, and 11 light fields from the Stanford dataset taken by a camera gantry~\cite{StanfordDataset}.
The Lytro Illum light fields are processed with the pipeline of Matysiak et al.~\cite{matysiak2018pipeline}.
For the synthetic light field dataset, all 28 light fields from the HCI benchmark~\cite{honauer2016benchmark} were used, as well as 160 light fields from the dataset of~\cite{alperovich2018light}.

For testing, 10 light fields are used: 2 from EPFL, 2 from INRIA, 2 from Stanford, and 4 from HCI.
All remaining light fields are used for training.

Test sparse light fields are sub-sampled from the original light fields with ratios $\alpha=2$ and $\alpha=4$.
More precisely, $9\times9$ views are extracted from input light fields and considered as dense ground truth, and $5\times5$ and $3\times3$ views are then sub-sampled to create sparse light fields.

We conduct the ablation experiments by comparing to several variants of the proposed framework.
First, we use the pre-trained model of SepConv as the baseline.
Since the dense light field ground truth is available, we fine-tuned the SepConv model using supervised training.
We also evaluate the influence of the cycle loss by training our framework using only the reconstruction loss.
We also assess the performance of our framework when vertical interpolation is performed before horizontal interpolation, as opposed applying horizontal interpolation first as shown in Figure~\ref{fig:overview}.

The numerical results are computed and averaged over all test light fields, and the comparison is presented in Table~\ref{table:ablation}.
As we can observe, our proposed method can outperform the pre-trained model even without the support of the ground-truth, and achieve competing performance compared to fully supervised fine-tuning.
It is also clear that the use of the cycle loss improves the performance of our framework.
In addition, we can see that cascading order of horizontal/vertical or vertical/horizontal interpolation has a non-negligible impact on the final performance.

\setlength{\tabcolsep}{3pt}
\begin{table}[!ht]
\begin{center}
\caption{\small{Quantitative results of the ablation study.}}
\begin{tabular}{ccccc}
\hline
\noalign{\smallskip}
\multirow{2}{*}{} & \multicolumn{2}{c}{$\alpha = 2$} & \multicolumn{2}{c}{$\alpha = 4$} \\
\cline{2-5}
\noalign{\smallskip}
 & PSNR(dB) & SSIM & PSNR(dB) & SSIM \\
\noalign{\smallskip}
\hline
\noalign{\smallskip}
SepConv Pretrained & 37.23 & 0.9880 & 34.66 & 0.9793\\
SepConv Supervised Fine-tuning & 38.40 & 0.9921 & 35.81 & 0.9831\\
Ours without Cycle Loss & 38.01 & 0.9883 & 35.25 & 0.9801\\
Ours with V-H CNN & 38.14 & 0.9889 & 35.67 & 0.9817\\
Ours Full Model & 38.30 & 0.9902 & 35.72 & 0.9830\\
\hline
\end{tabular}

\label{table:ablation}
\end{center}
\vspace{-0.4cm}
\end{table}
\setlength{\tabcolsep}{1.4pt}

\subsection{Comparison to Light Field View Synthesis Methods}

We use here the same test datasets as for the ablation study to compare the proposed framework against the pre-trained SepConv model, shearlet-based reconstruction~\cite{Shearlet}, and LFEPICNN~\cite{LFEPICNN_journal}.
We used the implementations provided by authors, and carefully selected their parameters to maximize their performance.

We show the quantitative results for each dataset separately in Table~\ref{table:lytro},~\ref{table:hci}, and~\ref{table:stanford}, as each dataset corresponds to a different disparity range.
Note that the results are averaged per dataset.

The shearlet-based reconstruction is almost always outperformed by all other methods, and while shearlet-based reconstruction and LFEPICNN are designed specifically for light fields, they are only competitive on the Lytro dataset which has a very narrow disparity range.
Our proposed framework consistently outperforms all other methods including SepConv.
In addition, our method is more robust when using sparser input datasets such as when using a sub-sampling ratio $\alpha=4$.

\setlength{\tabcolsep}{4pt}
\begin{table}[!ht]
\begin{center}
\caption{\small{Numerical results on the real-world Lytro datasets~\cite{Rerabek2016,INRIADataset}}}
\label{table:lytro}
\begin{tabular}{ccccc}
\hline
\noalign{\smallskip}
\multirow{2}{*}{} & \multicolumn{2}{c}{$\alpha = 2$} & \multicolumn{2}{c}{$\alpha = 4$} \\
\cline{2-5}
\noalign{\smallskip}
 & PSNR(dB) & SSIM & PSNR(dB) & SSIM \\
\noalign{\smallskip}
\hline
\noalign{\smallskip}
Shearlet & 33.10 & 0.9667 & 29.99 & 0.9361\\
LFEPICNN & 35.35 & 0.9864 & 32.06 & 0.9640\\
SepConv  & 35.30 & 0.9836 & 32.46 & 0.9712\\
Ours & \textbf{36.76} & \textbf{0.9876} & \textbf{33.62} & \textbf{0.9767}\\
\hline
\end{tabular}
\end{center}
\vspace{-0.4cm}
\end{table}
\setlength{\tabcolsep}{1.4pt}

\setlength{\tabcolsep}{4pt}
\begin{table}[!ht]
\begin{center}
\caption{Numerical results on the synthetic HCI dataset~\cite{honauer2016benchmark}}
\label{table:hci}
\begin{tabular}{ccccc}
\hline
\noalign{\smallskip}
\multirow{2}{*}{} & \multicolumn{2}{c}{$\alpha = 2$} & \multicolumn{2}{c}{$\alpha = 4$} \\
\cline{2-5}
\noalign{\smallskip}
 & PSNR(dB) & SSIM & PSNR(dB) & SSIM \\
\noalign{\smallskip}
\hline
\noalign{\smallskip}
Shearlet & 34.81 & 0.9734 & 29.88 & 0.8911\\
LFEPICNN & 34.25 & 0.9692 & 30.42 & 0.9172\\
SepConv  & 38.88 & 0.9943 & 36.23 & 0.9888\\
Ours & \textbf{39.87} & \textbf{0.9953} & \textbf{37.44} & \textbf{0.9913}\\
\hline
\end{tabular}
\end{center}
\vspace{-0.4cm}
\end{table}
\setlength{\tabcolsep}{1.4pt}

\begin{table}[!ht]
\begin{center}
\caption{Numerical results on the real-world Stanford Gantry datasets~\cite{StanfordDataset}}
\label{table:stanford}
\begin{tabular}{ccccc}
\hline
\noalign{\smallskip}
\multirow{2}{*}{} & \multicolumn{2}{c}{$\alpha = 2$} & \multicolumn{2}{c}{$\alpha = 4$} \\
\cline{2-5}
\noalign{\smallskip}
 & PSNR(dB) & SSIM & PSNR(dB) & SSIM \\
\noalign{\smallskip}
\hline
\noalign{\smallskip}
Shearlet & 31.44 & 0.8977 & 29.03 & 0.8484\\
LFEPICNN & 34.68 & 0.9407 & 30.46 & 0.8762\\
SepConv  & 37.80 & 0.9843 & 35.19 & 0.9788\\
Ours & \textbf{38.23} & \textbf{0.9853} & \textbf{36.47} & \textbf{0.9791}\\
\hline
\end{tabular}
\end{center}
\vspace{-0.4cm}
\end{table}
\setlength{\tabcolsep}{1.4pt}

\begin{figure}[!t]
\centering
\includegraphics[width=0.9\linewidth]{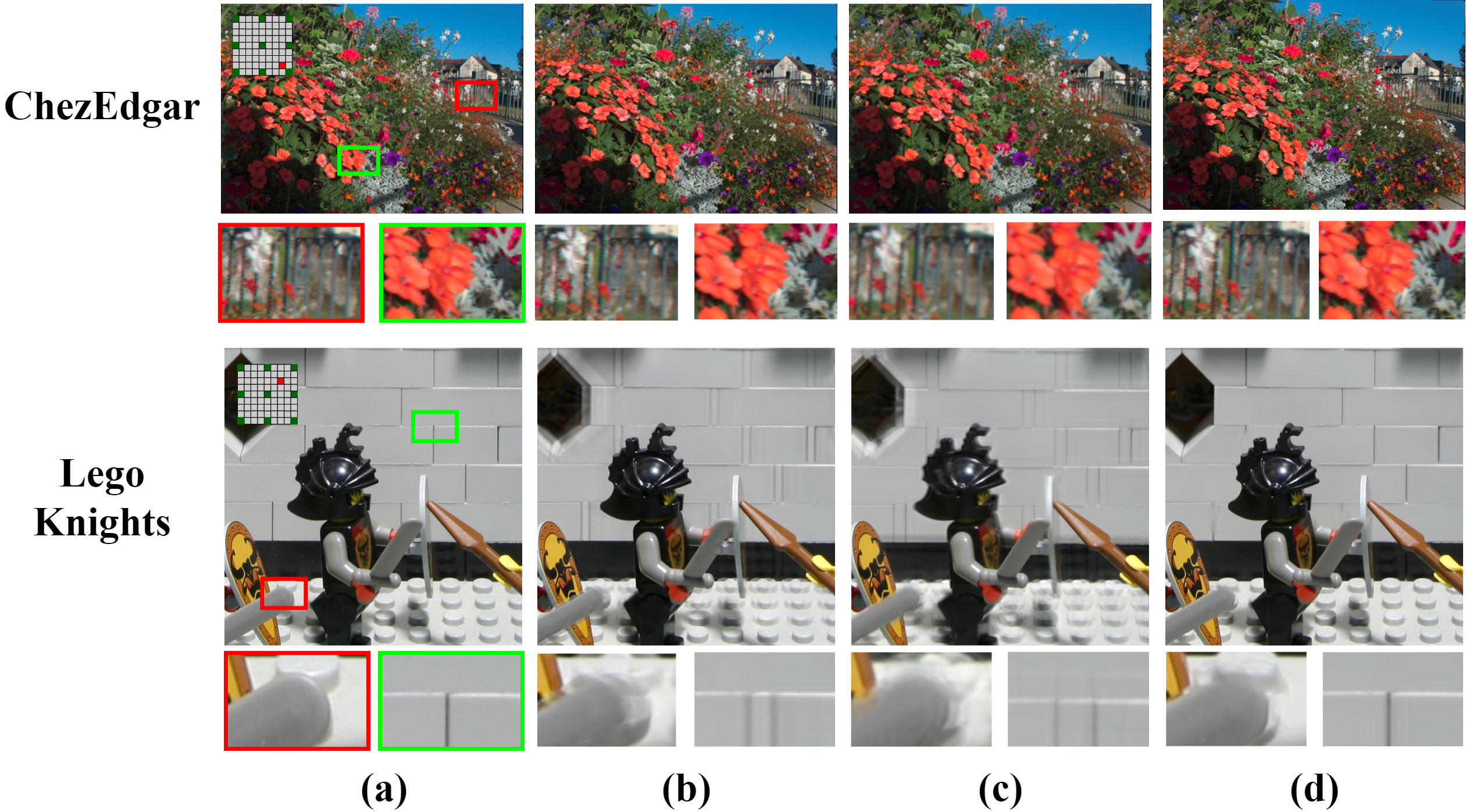}
\caption{\small{Visual comparison on the INRIA ChezEdgar and Stanford Lego Knights light fields. (a) Ground-truth. (b) Shearlet~\cite{Shearlet}. (c) LFEPICNN~\cite{LFEPICNN_journal}. (d) Ours}}
\label{fig:real-world}
\vspace{-0.5cm}
\end{figure}

We present visual comparisons for the \emph{ChezEdgar} and \emph{LegoKnights} light fields in Figure~\ref{fig:real-world}.
\emph{LegoKnights} is a challenging case as it has wider disparity than other test light fields and large texture-less regions.
Shearlet~\cite{Shearlet} and LFEPICNN~\cite{LFEPICNN_journal} both fail to produce plausible results and significant artifacts can be observed on challenging areas, such as the tip of the sword and bricks on the background wall.
In comparison, our proposed approach generates results closer to the ground-truth.
It demonstrates that our method is more robust to different real-world scenes and is able to produce more photo-realistic results for large disparity view synthesis.

\begin{figure}[!t]
\centering
\includegraphics[width=0.9\linewidth]{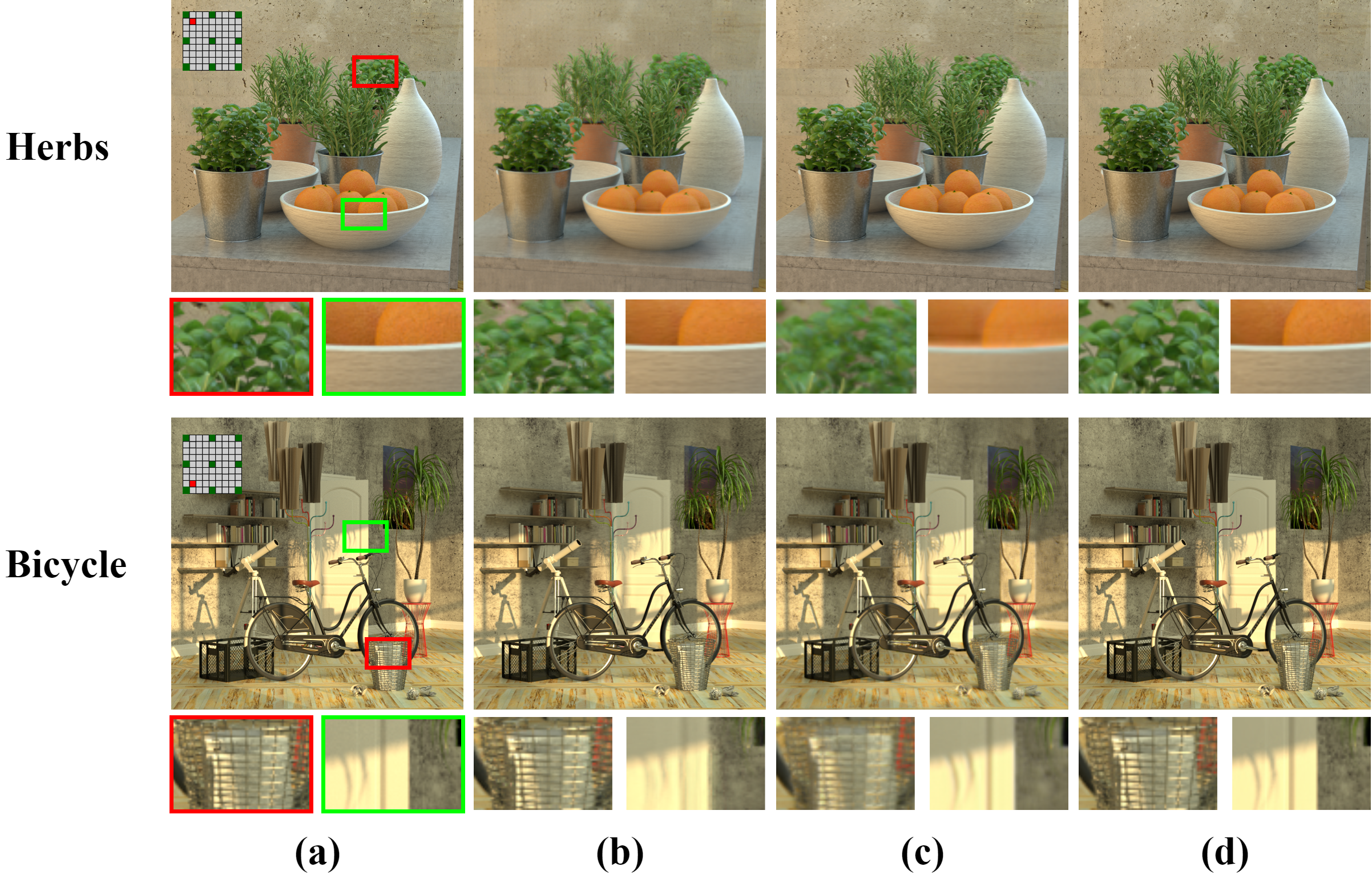}
\caption{\small{Visual comparison on the synthetic HCI dataset~\cite{honauer2016benchmark}. (a) Ground-truth. (b) Shearlet~\cite{Shearlet}. (c) LFEPICNN~\cite{LFEPICNN_journal}. (d) Ours}}
\label{fig:hci}
\vspace{-0.555cm}
\end{figure}

A visual comparison of synthetic scenes is presented in Figure~\ref{fig:hci} using \emph{Herbs} and \emph{Bicycle} from the HCI dataset~\cite{honauer2016benchmark}.
As we can observe, Shearlet~\cite{Shearlet} fails to reconstruct sharp details in texture-less regions, such as the door in \emph{Bicycle}.
The results of LFEPICNN~\cite{LFEPICNN_journal} are blurry in occluded regions, such as the leaves in \emph{Herbs} and the metal bin in \emph{Bicycle}.
Our method achieves best quantitative and qualitative performance on the synthetic HCI dataset, and shows robustness to occlusions and texture-less surfaces.

\section{Conclusions}
\label{sec:conclusion}
In this work, we proposed a novel self-supervised framework to reconstruct dense light fields by synthesizing novel intermediate light field views. 
To adopt small-sized light field datasets, we introduced the cycle consistency mechanism to fine-tune a pre-trained video interpolation method in a self-supervised fashion.
In this context, this method does not require paired ground-truth and is able to use for any low angular resolution light field input. 
The proposed method outperforms other state-of-the-art approaches on various light fields, in particular, given handling wide disparity inputs. 
In addition, our method can be adopted to any video interpolation approach, and let any 2D video interpolation into apply to light field data.
For future work, we may focus on adopting the proposed method to more challenging scenarios, such as very sparse light fields captured by camera arrays. 
This may require additional priors handle the sparsity.

\vspace{-0.155cm}

\bibliographystyle{IEEEtran}
\bibliography{conference_101719}

\clearpage

\end{document}